\documentclass{article}
\usepackage{amssymb,amsmath,graphicx}

\date{14 January 1997}

\newcommand{\gp}{\boldsymbol{p}}

\newcommand{\gr}{\boldsymbol{x}}
\newcommand{\gsig}{\boldsymbol{\sigma}}

\newcommand{\cE}{\mathcal{E}}

\ifx\begin{subarray}\undefined
  \newcommand{\sbst}{Sb}
\else
  \newcommand{\sbst}{subarray}
\fi

\title{Lower bound for the ground state energy of the no-pair Hamiltonian}

\author{C. Tix\thanks{E-mail: christti@math.uio.no} \\ Matematisk
  institutt, Universitetet i Oslo \\ Postboks 1053, N-0316 Oslo,
  Norway}

\begin{document}

\maketitle

\begin{abstract} 
  A lower bound for the ground state energy of a one particle
  relativistic Hamiltonian - sometimes called no-pair operator - is
  provided.
\end{abstract}

\section{Introduction}\label{s1} 

The multi-particle analogon of the operator
\begin{equation}\label{eq1}
  B = \Lambda_+ \left( D_0- \frac{e^2 Z}{|\gr|} \right)\Lambda_+
\end{equation}
is often used to describe relativistic effects in atoms and molecules,
see, e.g., \cite{IshikawaKoc1994,Jensenetal1996}.  Here $D_0$ is the
free Dirac operator $D_0= c\frac\hbar i \boldsymbol{\alpha}\cdot
\nabla + \boldsymbol{\beta} mc^2$ and
\begin{equation*}
  \Lambda_+(\gp) = \frac 12 \left(1+\frac{ c \boldsymbol{\alpha} \cdot
      \gp+ \boldsymbol{\beta} mc^2 }{E(\gp)}\right)
\end{equation*}
with $E(\gp)= (c^2\gp^2 + m^2 c^4)^{1/2}$ is the projection operator
on the electron subspace of the free Dirac operator in momentum space.
The underlying Hilbert space of $B$ is the space of square integrable
four spinors $\psi$ with $ \Lambda_+\psi=\psi$.

The operator $B$ was introduced for the two particle case by Brown and
Ravenhall~\cite{BrownRavenhall1951} (see
also~\cite{BetheSalpeter1957,Sucher1987}) to cure the continuum
dissolution of the two particle Dirac operator with Coulomb
interaction: since the Dirac operator is unbounded form below the two
particle Dirac operator has the whole real line as spectrum, the
eigenvalues of the one particle operator ``dissolve''. $B$ is called
no-pair operator or in the description of meson models (usually with
different potentials) ``reduced Salpeter
operator''~\cite{Olssonetal1996}. Recently, its multi particle
analogon was used in connection with the ``stability of relativistic
matter''--question~\cite{Liebetal1997}.

Similar to the Dirac operator with Coulomb potential, the no-pair
operator has a critical nuclear charge $Z_c$. Hardekopf and
Sucher~\cite{HardekopfSucher1984,HardekopfSucher1985} observed that
$Z_c=2/[(\pi/2 + 2/\pi)\alpha]$ with the fine structure constant
$\alpha:=e^2/(\hbar c)$ and investigated the ground state energy of
$B$ numerically. They claimed that, as for the Dirac equation, the
ground state energy vanishes for
$Z=Z_c$~\cite[p.~2025]{HardekopfSucher1985}. Evans et
al~\cite{Evansetal1996} proved that the energy $(\psi,B \psi)$ is
bounded from below by $ \alpha Z (1/\pi-\pi/4)mc^2$, if the nuclear
charge does not exceed the critical charge $Z_{c}$, otherwise it is
unbounded. Tix~\cite{Tix1997a} improved this bound to $mc^2\left(
  1-\alpha Z-0.002Z/Z_c \right)\ge 0.09$.  This last result shows the
difficulties to obtain accurate numerically results near the critical
coupling parameter and the need for sharp bounds. In this short
note we explore the results from~\cite{Tix1997a} by using some rather
trivial numerics to obtain a reliable lower bound for the ground state
energy for all $Z\le Z_c$.

\section{The Lower bound}\label{s2}

\begin{figure}[h]
  \begin{center}
    \includegraphics*[width=8cm]{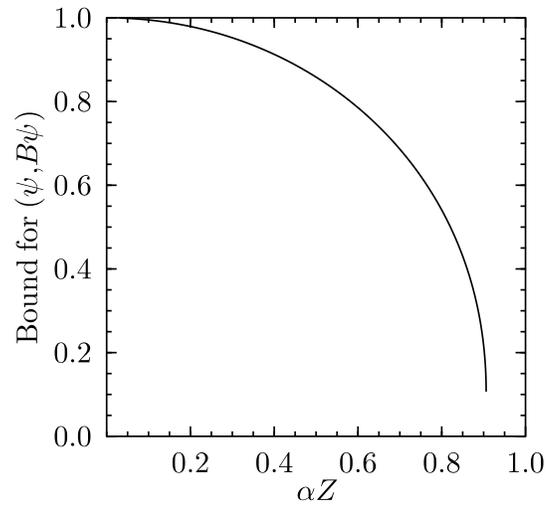}
    \caption{Lower bound for the energy $(\psi,B\psi)$ of the no-pair
      Hamiltonian} 
    \label{fig1}
  \end{center}
\end{figure}

A short summary of the method to obtain a lower bound for $(\psi,B
\psi)$ given in~\cite{Evansetal1996,Tix1997a} is provided.  Any
normalized four spinor in the electron subspace of $D_0$ can in be
written in momentum space as
\begin{equation}\label{2.2} 
  \widehat\psi(\gp) = \frac{1}{N(\gp)}
  \begin{pmatrix}
    \left(E_0+E(\gp)\right) u(\gp) \\
    c\gp\cdot\gsig u(\gp)
  \end{pmatrix}
\end{equation}
with the normalized Pauli spinor $u$, the Pauli-matrices $\gsig_j$ and
$E_0= E(\boldsymbol{0})$, $N(\gp)=[2E(\gp)(E(\gp)+E_0)]^{1/2}$.
In~\cite{Evansetal1996} it was shown that the minimizer of
$(\psi,B\psi)$ is radially symmetric, i.e., the minimizing $u$ is of
the form $u(\gp) = a(|\gp|)(\sqrt{4\pi} |\gp|)^{-1}(1,0)^*$ where $a$
is normalized and positive. This yields for the energy $(\psi,B\psi)$
\begin{equation}\label{partial}
  \int_0^\infty e(p)a(p)^2 d p -\frac{\alpha Z}{2\pi}
  \int_0^\infty\int_0^\infty a(p')k(p',p)a(p)d p d p'
\end{equation}
with  $e(p)=E(\gp)$, $p=|\gp|$, $k:=k_0+k_1$ and 
\begin{eqnarray}\label{2.8}
  k_j(p',p):=s_j(p')\,Q_j\!\left(\frac{p}{2p'} +
    \frac{p'}{2p}\right)s_j(p).
\end{eqnarray} 
The functions $Q_l$ are Legendre functions of the second kind
(see~\cite{Stegun1965} for the notation) and appear here for the same
reason as in the treatment of the non--relativistic hydrogen atom in
momentum space~\cite[problem 77]{Flugge1971} and $s_j(p)=\sqrt{1 +
  (-1)^j/e(p)}$.  Because of scaling we have assumed $\hbar=m=c=1$
in~\eqref{partial}.  Using the Schwarz inequality to introduce trial
functions $h_j(p)>0$ in the potential energy part of $(\psi,B\psi)$
\begin{multline*}\label{s10}
  \int_0^\infty\int_0^\infty a(p')k_j(p',p)a(p)d p d p' \\
  =\int_0^\infty \int_0^\infty a(p')s_j(p') \sqrt{Q_j}\,\,
  \sqrt{\frac{h_j(p)}{h_j(p')}}\,
  \sqrt{\frac{h_j(p')}{h_j(p)}}\,\sqrt{Q_j}\,\,
  a(p)s_j(p)\,  d p d p'\\
  \leq \int_0^\infty a(p)^2 s_j(p)^2 \int_0^\infty
  \frac{h_j(p')}{h_j(p)} Q_j\left( \frac{p}{2p'}+\frac{p'}{2p}\right)d
  p' \, dp
\end{multline*}
is obtained. Note that the operator $B$ is similar to the Herbst
operator $H=\sqrt{c^2 p^2 +m^2c^4}-e^2 Z/|\gr|$. The ground state
energy of $H$ is obtained by minimizing the quadratic
form~\eqref{partial} with $k=k_0$, without the square roots and with
$2Z$ instead of $Z$.  Since no error is introduced by using the
Schwarz inequality when $h_j/s_j$ is proportional to the ground state
wave function of~\eqref{partial}, this suggests a similar trial
function than the one that was used by Raynal et
al~\cite{Raynaletal1994} for the Herbst operator.  They obtained
excellent upper and lower bounds for the ground state of $H$.
Choosing the trial functions $h_j$ essentially as the Fourier
transform of $|\gr|^{\beta-1}e^{-\mu|\gr|}$, the lower bound
\begin{equation}
  \label{s18c}
  (\psi,B\psi) \ge \sup_{
    \begin{\sbst} \,
      -1<\beta <  1\\
      \quad \mu >0
    \end{\sbst}}  \,
  \inf_{p > 0} \cE[Z,\beta,\mu](p). 
\end{equation}
was obtained in~\cite{Tix1997a} where 
\begin{multline}
  \label{s18a}
  \cE[Z,\beta,\mu](p)=e(p)-\frac{\alpha Z}{2}
  \frac{\sqrt{p^2+\mu^2}}{\beta} \frac{\sin[ \beta\arctan (p/\mu)]}{
    \sin[(\beta+1) \arctan(p/\mu)]}
  \left(1+\frac{1}{e(p)}\right) \\
  - \frac{\alpha Z}{2}\left(1 - \frac{1}{e(p)} \right)
  \frac{\sqrt{p^2+\mu^2}}{\beta+2}\, \frac{P^{-3/2}_{\beta-1/2}
    \left(\mu (p^2+\mu^2)^{-1/2} \right)}{P^{-3/2}_{ \beta+1/2}
    \left(\mu (p^2+\mu^2)^{-1/2} \right) }
\end{multline}
with the associated Legendre functions $P^{-3/2}_{ \nu}$
(see~\cite[p. 1060]{Gradshteynryzhik1965}).
To cancel the large momentum terms in~\eqref{s18a}   $\beta\ge 0$ is
chosen to fulfill
\begin{equation}
  \label{eq31}
  \frac{2}{\alpha Z}= \frac{\tan(\beta\pi/2) }{\beta}+
  \frac{\beta}{(1-\beta^2)\tan(\pi\beta/2)}.
\end{equation}
The resulting function $\cE[Z,\beta,\mu]$ is numerically evaluated.
The values for the parameters $\beta$ and $\mu$ for some values of
$\alpha Z$ can be found in table~\ref{tab1} and a plot of the energy
over the coupling constant is shown in figure~\ref{fig1}.
\begin{table}[h]
  \begin{center}
    \leavevmode
    \begin{tabular}{|l|l|l|l|}\hline
      $\alpha Z$ &   $E$    & $\mu$ &$\beta$\\ \hline 
      0.9060   &  0.106   &  0.9843 &  0    \\ \hline
      0.9003   &   0.215  &  0.9526 &  0.1  \\ \hline
      0.8527   &  0.423   &  0.8652 &  0.3  \\ \hline
      0.7500   &  0.621   &  0.7395 &  0.5  \\ \hline
      0.5709   &  0.808   &  0.5552 &  0.7  \\ \hline
    \end{tabular}
    \caption{Lower bound for the energy  $E=\inf_p \cE
      [Z,\beta,\mu](p)$ and the corresponding parameters}
    \label{tab1}
  \end{center}
\end{table}

\clearpage

\section{Acknowledgment}
\label{sec4}

This work was partially supported by the European Union under grant
ERB4001GT950214 and under the TMR-network grant FMRX-CT 96-0001.


\begin{thebibliography}{10}

\bibitem{BetheSalpeter1957}
Hans~A. Bethe and Edwin~E. Salpeter.
\newblock Quantum mechanics of one- and two-electron atoms.
\newblock In S.~Fl{\"u}gge, editor, {\em Handbuch der {P}hysik, {XXXV}}, pages
  88--436. Springer, Berlin, 1 edition, 1957.

\bibitem{BrownRavenhall1951}
G.E. Brown and D.G. Ravenhall.
\newblock {\em Proc. Roy. Soc. London A}, 208(A 1095):552--559, September 1951.

\bibitem{Evansetal1996}
William~Desmond Evans, Peter Perry, and Heinz Siedentop.
\newblock {\em Commun. Math. Phys.}, 178(3):733--746, July 1996.

\bibitem{Flugge1971}
Siegfried Fl{\"u}gge.
\newblock {\em Practical Quantum Mechanics I}, volume 177 of {\em
  Grund\-leh\-ren der mathematischen Wissenschaften}.
\newblock Springer-Verlag, Berlin, 1 edition, 1982.

\bibitem{Gradshteynryzhik1965}
I.S. Gradshteyn and I.M. Ryzhik.
\newblock {\em Table of integrals, series, and products}.
\newblock Academic Press, New York and London, 4 edition, 1965.

\bibitem{HardekopfSucher1984}
G.~Hardekopf and J.~Sucher.
\newblock {\em Phys. Rev. A}, 30(2):703--711, August 1984.

\bibitem{HardekopfSucher1985}
G.~Hardekopf and J.~Sucher.
\newblock {\em Phys. Rev. A}, 31(4):2020--2029, April 1985.

\bibitem{IshikawaKoc1994}
Y.~Ishikawa and K.~Koc.
\newblock {\em Phys.~Rev.~A}, 50(6):4733--4742, December 1994.

\bibitem{Jensenetal1996}
Hans J{\"o}rgen~Aa. Jensen, Kenneth~G. Dyall, Trond Saue, and Knut~F{\ae}gri
  Jr.
\newblock {\em J. Chem. Physics}, 104(11):4083--4097, March 1996.

\bibitem{Liebetal1997}
Elliott~H. Lieb, Heinz Siedentop, and Jan~Philip Solovej.
\newblock Stability and instability of relativistic electrons in classical
  electromagnetic fields.
\newblock {\em J. Stat. Phys.}, submitted, 1997.

\bibitem{Olssonetal1996}
M.~G. Olsson, S.~Veseli, and K.~Williams.
\newblock {\em Phys. Rev. D}, 53(1):504--513, 1996.

\bibitem{Raynaletal1994}
J.~C. Raynal, S.M. Roy, V.~Singh, A.~Martin, and J.~Stubbe.
\newblock {\em Phys. Lett. B}, 320:105--109, 1994.

\bibitem{Stegun1965}
Irene~A. Stegun.
\newblock Legendre functions.
\newblock In Milton Abramowitz and Irene~A. Stegun, editors, {\em Handbook of
  Mathematical Functions with Formulas, Graphs, and Mathematical Tables},
  chapter~8, pages 331--353. Dover Publications, New York, 1965.

\bibitem{Sucher1987}
J.~Sucher.
\newblock {\em Phys. Scripta}, 36:271--281, 1987.

\bibitem{Tix1997a}
C.~Tix.
\newblock Strict positivity of a relativistic hamiltonian due to Brown and
  Ravenhall.
\newblock {\em Bull. London Math. Soc.}, submitted, 1997.

\end{thebibliography}
\end{document}